\documentclass{article}

\usepackage{PRIMEarxiv}
\usepackage{booktabs}
\usepackage[utf8]{inputenc} %
\usepackage[T1]{fontenc}    %
\usepackage{hyperref}       %
\usepackage{url}            %
\usepackage{booktabs}       %
\usepackage{amsfonts}       %
\usepackage{nicefrac}       %
\usepackage{microtype}      %
\usepackage{lipsum}
\usepackage{fancyhdr}       %
\usepackage{graphicx}       %
\usepackage{subcaption}
\usepackage{soul}
\usepackage{caption}       %
\graphicspath{{figures/}}     %
\usepackage[table,xcdraw]{xcolor}
\usepackage{array}
\usepackage{colortbl}
\usepackage{placeins}
\usepackage{soul,color}
\usepackage{booktabs} 
\usepackage{amsmath}
\usepackage{makecell}
\usepackage{float}  %
\title{Foundation-Model Surrogates Enable Data-Efficient Active Learning for Materials Discovery
\thanks{\textit{\underline{Citation}}: 
\textbf{J.Hu et al. Foundation model based In-context active learning for accelerated materials discovery. 15 Pages.... DOI:000000/11111.}} 
}

\author{%
  Jeffrey Hu$^1$, Rongzhi Dong$^2$,Ying Feng$^3$, Ming Hu $^4$, Jianjun Hu $^2$\thanks{Corresponding author: jianjunh@cse.sc.edu}\\
  $^1$Department of Materials Science and Engineering, University of Illinois Urbana Champaign, Champaign,IL, USA\\
  $^2$Department of Computer Science \& Engineering, University of South Carolina, Columbia, SC, USA\\
  $^3$Zuoyue Honors College, Hangzhou Dianzhi University, Hangzhou, China\\ 
  $^4$Department of Mechanical Engineering, University of South Carolina, Columbia, SC, USA\\
  }

\begin{document}
\maketitle

\begin{abstract}

Active learning has emerged as a powerful strategy for accelerating materials discovery by guiding expensive synthesis and characterization experiments toward the most informative candidates. However, its effectiveness critically depends on the surrogate model used to estimate both predictions and uncertainty. In experimental materials science, datasets are typically extremely small—often containing only tens to hundreds of samples—making reliable surrogate modeling challenging. Gaussian processes provide principled uncertainty estimates but often underfit complex composition–property relationships, while high-capacity models such as random forests and neural networks can capture nonlinear relationships but frequently produce unreliable uncertainty estimates in small-data regimes.

Here we introduce In-Context Active Learning (ICAL), a framework that replaces conventional surrogate models with a foundation model trained to approximate Bayesian inference over tabular data. By leveraging a meta-learned prior derived from millions of synthetic regression tasks, the model performs probabilistic inference in a single forward pass without dataset-specific retraining, enabling expressive prediction and calibrated uncertainty in small-data settings.

We evaluate ICAL across ten benchmark datasets spanning copper alloys, bulk metallic glasses, crystal lattice thermal conductivity, and electrolyte materials. ICAL consistently reduces the number of evaluations required to identify optimal materials, achieving a mean reduction of 52\% relative to Gaussian-process-based active learning and 29.77\% relative to random-forest baselines. 
Mechanistic analysis shows that these improvements arise from superior uncertainty calibration, allowing acquisition functions to more effectively balance exploration and exploitation during the discovery process.

These results demonstrate that foundation models can serve as effective surrogate models for data-efficient active learning, suggesting a new paradigm for accelerating discovery in small-data experimental sciences such as materials science, chemistry, biology, and other experimental disciplines characterized by small datasets and expensive evaluations.

\end{abstract}


\section{Introduction}

The discovery and optimization of new materials is a central challenge in science and engineering, with far-reaching implications for energy storage, structural alloys, catalysis, semiconductors, and drug design \cite{lookman2019active,lookman2026materials,chen2026survey,omidvar2024accelerated,wang2023accelerated,moon2024active,noh2024integrated}. Traditional Edisonian trial-and-error experimentation is both time-consuming and extraordinarily expensive. The cost of synthesizing and characterizing a single alloy or crystal material — including arc melting, X-ray diffraction, electron microscopy, and mechanical or functional property testing, with labor — ranges from approximately \$500 to \$3{,}000 for structural alloys and can reach \$20{,}000 or more for advanced functional materials. These costs create an essential bottleneck: the compositional and processing spaces of modern materials are combinatorially vast, yet only a tiny fraction can be explored experimentally. Active learning (AL), especially Bayesian Optimization (BO), has emerged as a powerful paradigm to address this challenge by coupling machine learning (ML) predictions with adaptive data acquisition, iteratively steering experiments toward the most informative or highest-performing candidates and thereby reducing the number of costly evaluations required to reach a design target \cite{lookman2019active, settles2009active, chen2026survey,ding2025leveraging}.
 
Since the earliest active learning materials discovery studies in 2016 \cite{xue2016accelerated}, the breadth of AL's impact as a computation engine driving materials discovery is substantial and growing. In functional materials, Yuan et al.\ demonstrated accelerated discovery of large electrostrains in BaTiO$_3$-based piezoelectrics \cite{yuan2018accelerated}, and Bassman et al.\ applied AL to layered materials design \cite{bassman2018active}. In electrocatalysis, Moon et al.\ used AL to discover a champion four-metal perovskite oxide for oxygen evolution \cite{moon2024active}, Suvarna et al.\ streamlined development of high-performance catalysts for higher alcohol synthesis \cite{suvarna2024active}, and Mok et al.\ identified electrocatalysts for CO$_2$ reduction \cite{ding2025leveraging}. For alloy systems, Kim et al.\ searched for optimal multi-metallic alloy catalysts \cite{kim2022searching}, and Rao et al.\ developed closed-loop AL frameworks for high-entropy Invar alloys \cite{rao2022machine}. Beyond metals, Kunkel et al.\ applied AL to organic semiconductor discovery \cite{kunkel2021active}. AL has also transformed chemistry through self-driving laboratories (SDLs), which automate experimental planning and execution to accelerate the research pipeline in both chemistry and materials science \cite{tom2024self, canty2025science, szymanski2023autonomous, Abolhasani2023}. In drug design, Bayesian AL frameworks have been applied to drug screening \cite{tosh2025bayesian} and regression-based AL has been used to accelerate large-scale virtual library docking \cite{Marin2024}. Across all these domains, the common thread is that AL enables order-of-magnitude reductions in the number of experiments required relative to conventional design-of-experiments or random search \cite{jang2025active,lookman2019active,kusne2020fly,ma2025active,kotthoff2021bayesian}.
 
\subsection*{Current Practice and Limitations of Surrogate Models in Materials Active Learning}
 
Despite its broad success, AL in materials science is heavily dependent on the choice of surrogate model, and current practice reveals significant weaknesses in all widely used approaches. A recent comprehensive survey \cite{chen2026survey} and benchmark study \cite{liang2021benchmarking} show that Gaussian Process (GP) regression has historically been the dominant surrogate model in Bayesian Optimization (BO)-based AL, valued for its native probabilistic output and its principled uncertainty estimates that directly enable the exploration–exploitation trade-off inherent to acquisition functions such as Expected Improvement (EI) and Upper Confidence Bound (UCB) \cite{raihan2024accelerating, lookman2019active}. GP-based BO has been applied extensively: to catalyst design \cite{suvarna2024active, ji2024multi}, alloy discovery \cite{rao2024active, wang2022accelerated, lee2025active}, and shape memory alloys \cite{lookman2019active}, among many others. However, GP surrogates suffer from a fundamental modeling limitation: their rigid parametric assumptions about the feature–property landscape lead to high bias, making them prone to underfitting complex, nonlinear, or high-dimensional composition–property relationships \cite{alvi2024hierarchical, chen2026survey}. Furthermore, standard GPs scale cubically with data size, limiting their applicability as datasets grow \cite{alvi2025deep}.
 
Random Forest (RF) models have been proposed as a practical alternative, offering greater model flexibility, lower sensitivity to hyperparameter choice, and no strict distributional assumptions \cite{liang2021benchmarking}. Benchmark studies confirm that RF and GP with anisotropic kernels offer comparable top-tier performance, with RF being advantageous in terms of computational cost and ease of use \cite{liang2021benchmarking}. RF has therefore been adopted widely as a surrogate in AL for materials design too\cite{rao2022machine, johnson2024active, trehern2022data, cao2026bgolearn}. However, RF's high model capacity exposes it to a different failure mode: in the small-data regimes that are characteristic of experimental materials discovery \cite{liu2023data}, RF is prone to overfitting, and its uncertainty estimates — derived from ensemble variance across decision trees — are heuristic rather than principled. This makes RF uncertainty unreliable as a guide for query selection, particularly in the early iterations of an AL campaign when the training set is smallest and reliable uncertainty quantification is most critical \cite{dale2025when}.
 
Neural network (NN) surrogates, including multi-layer perceptrons (MLPs), deep ensembles, and graph neural networks such as ALIGNN, share RF's high model capacity but are arguably even more susceptible to overfitting in low-data regimes due to their larger number of parameters and the absence of RF's built-in regularization through bagging and feature subsampling \cite{rao2022machine, dale2025when, deshmukh2022active}. Ensemble-based uncertainty quantification for NNs — whether through deep ensembles, Monte Carlo dropout, or bootstrapping — introduces additional computational cost and remains heuristic in nature. Deep Gaussian Processes (DGPs) have been proposed to overcome the limited expressiveness of standard GPs by stacking GP layers to model complex hierarchical feature relationships \cite{alvi2025deep}, while hierarchical GP frameworks have been developed to handle high-entropy alloy composition spaces \cite{alvi2024hierarchical}. Despite these advances, the fundamental dilemma remains: GP models have reliable uncertainty but insufficient modeling capacity; RF and NN models have high capacity but unreliable uncertainty in small-data regimes. This tension between model fidelity and uncertainty quality represents a core, unresolved challenge in AL-based materials discovery. In addition, the data scarcity widely exist in materials discovery. A recent survey \cite{liu2023data} over 109 data sets, shows that $\sim$57\% have <500 samples, $\sim$67\% comprise <1000 samples and only $\sim$21\% of data sets contain >2000 samples. The small datasets inherent in materials science has posed significant challenges to machine learning and active learning \cite{zhang2018strategy,xu2023small,zhou2025machine}

 This tension between predictive capacity and uncertainty reliability represents a central challenge in active-learning-driven materials discovery. We hypothesize that foundation models trained to approximate Bayesian inference over tabular data can serve as universal surrogate models for small-data experimental discovery, resolving this long-standing trade-off and enabling more efficient experimental exploration.

\begin{figure}[ht!]
  \centering
  \includegraphics[width=0.9\linewidth]{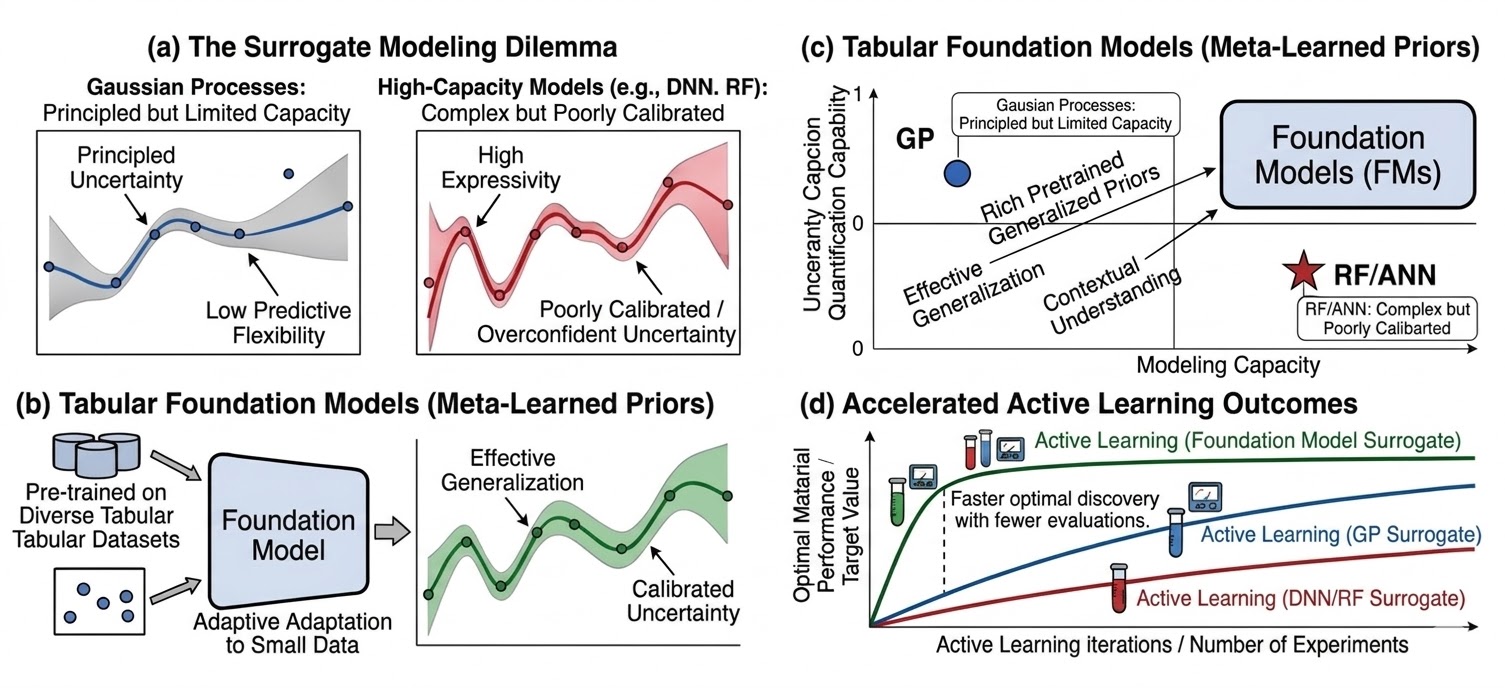}
  \caption{Foundation models resolve the surrogate modeling dilemma in active learning.
(a) Conventional surrogate models used in active learning face a trade-off between predictive capacity and reliable uncertainty estimation. Gaussian processes provide principled uncertainty but limited modeling flexibility, while high-capacity models such as random forests and neural networks can capture complex relationships but often produce poorly calibrated uncertainty in small-data regimes. (b) Foundation models trained to approximate Bayesian inference over tabular data combine expressive prediction with calibrated uncertainty through meta-learned priors. (c) Active learning with foundation-model surrogates identifies achieve good balance of modeling capacity and uncertainty estimation capability. (d) FM-based AL can find optimal materials with fewer experimental evaluations.}
  \label{fig:concept}
\end{figure}

\subsection*{Motivation: Foundation Model Surrogate for Active Learning}

To address this limitation, we propose In-Context Active Learning (ICAL), a framework that replaces conventional surrogate models with a foundation model trained to approximate Bayesian inference over tabular data  (Figure \ref{fig:concept}). Unlike traditional surrogates that must be trained from scratch for each dataset, the model leverages a meta-learned prior derived from large-scale synthetic training tasks, enabling expressive prediction and calibrated uncertainty even in extremely small-data regimes typical of experimental discovery.

Our ICAL algorithm uses TabPFN as the surrogate model, which is a transformer-based foundation model that has been meta-trained via in-context learning on a large corpus of synthetic tabular datasets to approximate Bayesian inference over small tabular data problems \cite{hollmann2025tabpfn}. Unlike GP, RF, and NN surrogates that are fit from scratch on each new dataset, TabPFN leverages a pre-trained prior over functions and performs inference in a single forward pass without any gradient-based optimization on the target dataset. Critically, TabPFN outputs a full predictive distribution — not merely a point estimate with a heuristic variance proxy — because it was explicitly meta-trained to represent posterior uncertainty over tabular inputs. This makes it a principled probabilistic surrogate that is both expressive enough to capture complex composition–property relationships and capable of producing reliable uncertainty estimates in the small-data, high-dimensional settings that characterize experimental materials discovery.
 
We argue that these properties make TabPFN uniquely suited as a surrogate model for AL in materials science, which has demonstrated their high performance in small data materials property prediction \cite{li2025context}. The predominant regimes of interest — early-stage alloy design, catalyst screening, and crystal property optimization — involve datasets of tens to a few hundred labeled examples, precisely the regime where TabPFN's meta-trained in-context inference provides the greatest advantage over both the underfitting of GP and the overfit-prone, heuristically uncertain RF and NN alternatives. Moreover, in-context learning enables TabPFN to rapidly adapt to new materials tasks without retraining, enabling seamless integration into the iterative AL loop. In this work, we propose and evaluate an In-Context Bayesian Active Learning framework that replaces conventional GP and ensemble surrogate models with TabPFN, and we additionally introduce an improved sample selection strategy that balances both quality (acquisition function score) and diversity to mitigate redundant querying — a known failure mode of greedy acquisition in materials spaces.

In this work, we propose and evaluate the In-Context Active Learning (ICAL) algorithm, which replaces conventional GP and RF surrogate models with TabPFN --- a transformer-based foundation model meta-trained to approximate Bayesian inference over tabular data --- for pool-based materials discovery. We benchmark ICAL against GP and RF baselines across 10 materials datasets spanning copper alloys, bulk metallic glasses, and crystal lattice thermal conductivity, covering diverse property targets and feature representations. Our main findings are as follows. First, TabPFN achieves the lowest number of extra evaluations on 8 out of 10 datasets, with a mean saving of 52\% relative to GP and 29.77\% relative to RF averaged across all TabPFN-winning datasets, demonstrating consistent superiority across diverse material classes and property targets. Second, TabPFN's advantage is mechanistically grounded in superior uncertainty quantification: cross-validation experiments on the Cu-alloy electrical conductivity and LTC datasets confirm that TabPFN achieves the lowest Negative Log-Likelihood and Area Under the Sparsification Error curve among all three surrogates, confirming that its in-context posterior produces well-calibrated, informative uncertainty estimates that guide acquisition functions more efficiently than GP's over-conservative intervals or RF's heuristic ensemble variance. Third, we identify a critical minimum data threshold effect: TabPFN requires approximately 10--20\% of the candidate pool as initial training data before its modeling advantage fully materializes, below which GP may perform comparably or better. Fourth, feature representation is a decisive factor in active learning performance that is largely independent of regression accuracy: high-dimensional Magpie features ($>$120 descriptors) consistently degrade AL performance for copper alloy and LTC datasets due to overfitting in the surrogate's uncertainty estimation, while the same Magpie features yield the best AL performance for glass-forming alloy discovery where the more complex composition--property landscape benefits from richer physicochemical descriptors. Finally, we introduce normalized and hybrid acquisition functions that address the scale instability of raw acquisition scores in small-data regimes, further improving TabPFN's active learning efficiency across datasets.

We hypothesize that foundation models trained to approximate Bayesian inference over tabular data can serve as universal surrogate models for small-data experimental discovery, resolving the long-standing trade-off between predictive capacity and reliable uncertainty estimation that limits conventional active learning approaches. This limitation arises from the small-data nature of experimental science. Foundation models offer a potential solution to this problem by providing meta-learned priors over functions.

\section{Method}
\label{sec:headings}

\subsection{Pool based materials discovery using active learning}

Active learning-based materials design has been widely used in materials discovery \cite{lookman2019active,chen2020generative,akbari2025machine}. These approaches can be broadly classified into two categories. The first is open-ended materials discovery, in which new samples are generated based on a machine learning model. The second category focuses on exploring a fixed set of candidate materials to screen for those with desired properties. While open-ended active learning design fits many realistic scenarios, it has the drawback of relying on real experimental validation or computational simulation (such as DFT) to obtain ground truth property values for evaluating algorithm performance. In contrast, using a fixed set of materials with annotated properties allows for a more direct evaluation of active learning performance. This is achieved by starting with a small proportion of samples as the initial training set and treating the remainder as unannotated. The goal is then to determine how many samples must be evaluated to find the target sample with the best property.

\begin{figure}[ht!]
  \centering
  \includegraphics[width=0.9\linewidth]{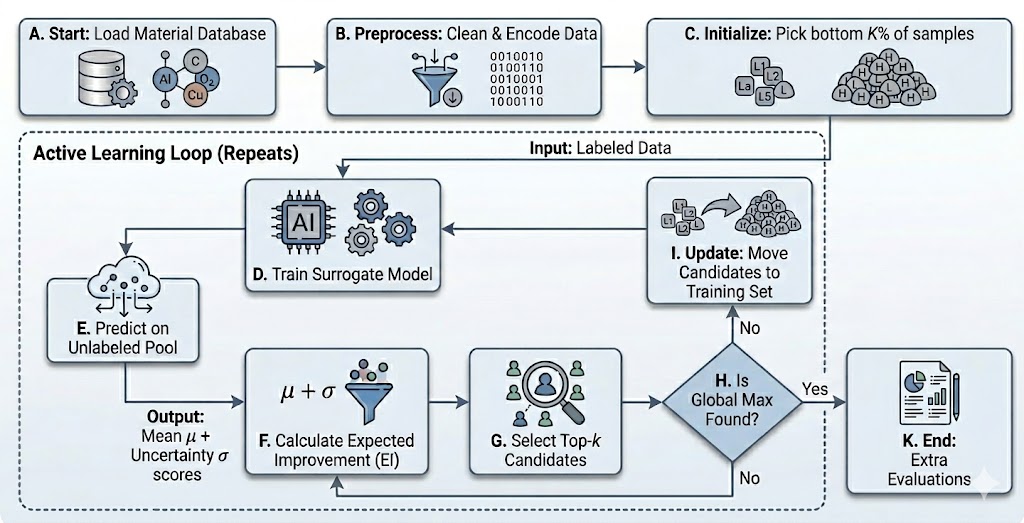}
  \caption{Pool-based active learning pipeline for discovering new materials. F: EI can be replaced with any other acquisition functions. G: k=1 is used for maximum sample efficiency. H: Top-1 task can be replaced with discovering Any of All of Top-K tasks.}
  \label{fig:framework}
\end{figure}

Our study adopts this second category of active learning-based screening for materials discovery or pool-based active learning, following the active learning materials design benchmark study in \cite{liang2021benchmarking} . The minor difference is that our initial population is selected as a given percentage of samples with the lowest/highest properties instead of random selection to reduce the experiment uncertainty. The main framework is illustrated in Figure \ref{fig:framework}. The process begins with data ingestion and preparation, where a raw materials database is loaded, cleaned, and encoded to convert categorical features into a numerical format suitable for machine learning. The process initializes by randomly selecting a small percentage of these materials to form the initial ``labeled'' training set, while the remaining majority constitutes the unlabeled pool.
The core of the methodology is the iterative active learning loop. In each cycle, a surrogate AI model is trained using the current labeled data. This model is then tasked with predicting the properties of the entire unlabeled pool, outputting both a predicted mean value ($\mu$) and an associated uncertainty ($\sigma$) for each candidate. These outputs are combined to calculate the Expected Improvement (EI), a strategic metric that balances exploration (high uncertainty) and exploitation (high predicted mean). The top-$k$ candidates with the highest EI scores are selected for evaluation. A decision step determines if the global maximum property has been identified among these candidates. If the maximum has not been found, the selected candidates are moved from the unlabeled pool into the training set—simulating the acquisition of new experimental labels—and the model is retrained in the next iteration. Once the global maximum is successfully identified, the loop terminates, and the total number of evaluations required to find the optimal solution is reported. In AL, each surrogate model may be configured with different acquisition functions. We have evaluated six acquisition functions as different ML models may work best with different acquisition functions (Table \ref{tab:acquisition_functions}).

\FloatBarrier

\begin{table}[htbp]
    \centering
    \caption{Summary of Active Learning Acquisition Functions}
    \label{tab:acquisition_functions}
    \renewcommand{\arraystretch}{1.5}
    \begin{tabular}{|l|c|p{8cm}|}
        \hline
        \textbf{Acquisition Strategy} & \textbf{Mathematical Formulation} & \textbf{Description \& Scaling Notes} \\
        \hline
        Standard UCB & $\mu(x) + \beta\sigma(x)$ & Balances mean and uncertainty using raw, \textbf{unnormalized} values (prone to scale issues). \\
        \hline
        Normalized UCB & $\mu_{norm}(x) + \beta\sigma_{norm}(x)$ & \textbf{Scale-invariant.} Both $\mu$ and $\sigma$ are normalized to $[0, 1]$ before scaling with $\beta$. \\
        \hline
        Hybrid UCB & $\mu_{norm}(x) + \beta\sigma_{GP\_norm}(x)$ & Uses normalized mean from TabPFN and normalized uncertainty from a Gaussian Process. \\
        \hline
        Standard EI & $(\mu(x) - f_{best})\Phi(Z) + \sigma(x)\phi(Z)$ & Calculates Expected Improvement using raw units without a trade-off parameter. \\
        \hline
        Normalized EI & $(\hat{\mu}(x) - \hat{f}_{best} - \xi)\Phi(Z) + \hat{\sigma}(x)\phi(Z)$ & Calculates Expected Improvement using min-max normalized predictions and a trade-off parameter $\xi$, where $Z = \frac{\hat{\mu}(x) - \hat{f}_{best} - \xi}{\hat{\sigma}(x)}$.\\
        \hline
        Hybrid EI & Same as Normalized EI & Computes Normalized EI $\xi$ utilizing the mean from TabPFN and the uncertainty from the GP. \\
        \hline
    \end{tabular}
    
    \vspace{2mm}
    \begin{flushleft}
    \footnotesize
    \textbf{Note:} 
    $\mu(x), \sigma(x)$: Predicted mean and standard deviation (uncertainty) for candidate material $x$.\\
    $\mu_{norm}, \sigma_{norm}$: Min-max normalized mean and standard deviation, scaled strictly to $[0, 1]$.\\
    $\beta$: Static exploration parameter dictating the trade-off between exploiting the mean and exploring uncertainty.\\
    $f_{best}$: The maximum target property value discovered in the dataset so far.\\
    $\xi$: A minimum threshold or trade-off parameter defining the required margin of improvement over $f_{best}$.\\
    $Z$: The standardized dimensionless improvement score, defined as $Z = (\mu_{norm}(x) - f_{best\_norm} - \xi) / \sigma_{norm}(x)$.\\
    $\Phi(\cdot), \phi(\cdot)$: The Cumulative Distribution Function (CDF) and Probability Density Function (PDF) of the standard normal distribution.\\
    $\mathcal{U}, \mathcal{N}$: Uniform and Normal probability distributions.
    \end{flushleft}
\end{table}

\subsection{In context active learning (ICAL) for materials discovery}

Recent literature reviews find that currently the mainstream practices of active learning based materials discovery are based on Bayesian optimization with conventional machine learning surrogate models such as Gaussian Process, Random Forest,  Neural networks, and gradient boosting trees \cite{xu2025applications,liang2021benchmarking,chen2026survey,wang2022accelerated}. However, these surrogate models either suffer from their inherent bias and lack of complex modeling capacity or from their overfitting risk in active learning based materials discovery. 

We propose the In-context Active Learning (ICAL) algorithm to address these issues in current active learning-driven materials discovery (ALMD) using TabPFN, an in-context learning foundation model, as a novel surrogate model. We benchmark it against the most widely adopted surrogate models in active learning practice, including Gaussian Processes (GP) and Random Forest (RF). Our ICAL algorithm leverages three distinctive modeling capacities of TabPFN to achieve superior active learning performance: (1) significantly stronger modeling capability for small datasets compared to GP, mitigating the underfitting risk arising from rigid parametric assumptions; (2) lower overfitting risk compared to RF (and MLP), attributed to its meta-learned inductive biases and its function regression from large-scale pretraining over synthetic regression datasets rather than relying solely on the scarce labeled data available in ALMD; and (3) native probabilistic regression capability, eliminating the need for computationally expensive ensemble methods or approximate Bayesian inference techniques such as Monto Carlo (MC) Dropout. 

\paragraph{Normalized EI/UCB}: We also recognize a fundamental issue in materials active learning. The first one is the scale issue in acquisition functions when using mean and standard deviation to calculate ranking scores, due to the extremely high variance in out-of-distribution predictions by surrogate models trained with small datasets. It is difficult to balance the mean and uncertainty. We
propose to normalize both the mean and standard deviation before ranking score calculation, leading to our normalized UCB and normalized EI acquisition function (See Table \ref{tab:acquisition_functions}). While this normalization may disrupt the theoretical rigor of original GP model based BO, our experiments showed that it leads to better performance for many datasets. It also brings the benefit of easy across-dataset hyper-parameter setting for e.g. the beta parameter in UCB. 

\paragraph{Hybrid Acquisition Functions:} We also propose hybrid acquisition functions (See Table \ref{tab:acquisition_functions}) to exploit the high modeling capacity of Tabpfn and RF and GP's geometry-sensitive uncertainty quantification. While TabPFN demonstrates superior predictive accuracy as a surrogate model, its uncertainty quantification relies on in-context probabilistic estimates that may not fully capture the geometric structure of the feature space. Gaussian Processes, by contrast, produce uncertainty estimates grounded in kernel-based covariance functions that explicitly encode spatial correlations between data points — a property particularly valuable in materials discovery, where property landscapes often exhibit smooth, structured variation across compositional or configurational spaces. We therefore propose a hybrid acquisition function combining UCB and EI that pairs TabPFN's mean prediction with GP's kernel-derived variance estimation. This decoupling exploits the complementary strengths of both models: TabPFN contributes a more accurate and flexible predictive mean, especially in low-data regimes, while GP contributes geometrically informed uncertainty estimates that better reflect the local density and distribution of the labeled sample pool. The resulting hybrid acquisition strategy achieves a more effective exploration-exploitation balance, as the GP variance naturally encourages sampling in structurally underexplored regions of the feature space that TabPFN's uncertainty estimates alone may overlook. This explains the observed performance gains on datasets whose property landscapes exhibit strong spatial correlation structure, while also clarifying why the improvement is dataset-dependent — on datasets with irregular or discontinuous property landscapes, the geometric assumptions embedded in GP's covariance structure may offer less advantage.

To justify our ICAL active learning algorithm, we describe the characteristics of TabPFN and compare it to two baseline models GP and RF. Table \ref{tab:surrogate_models} summarize how these models calculate their mean and uncertainty. 

\begin{table*}[htbp]
    \label{tab:mean_std}
    \centering
    \renewcommand{\arraystretch}{1.6}
    \caption{Summary of Predictive Mean and Uncertainty Estimation by Surrogate Models}
    \label{tab:surrogate_models}
    \begin{tabular}{p{3cm} p{6.5cm} p{6.5cm}}
        \toprule
        \textbf{Surrogate Model} & \textbf{Predictive Mean} ($\mu(x)$) & \textbf{Predictive Uncertainty} ($\sigma(x)$) \\
        \midrule
        
        \textbf{TabPFN} & 
        The median (50th percentile) of the output distribution predicted directly by the pre-trained transformer network. & 
        Derived from the predicted 95\% confidence interval. Calculated as the difference between the 97.5th and 2.5th predicted percentiles divided by $3.92$. \\
        
        \raggedright\arraybackslash \textbf{Random Forest (RF)} & 
        The average prediction across all $T$ individual decision trees in the ensemble: \newline
        $\mu(x) = \frac{1}{T} \sum_{t=1}^T \hat{y}_t(x)$ & 
        The standard deviation of predictions across the ensemble of trees, serving as a measure of epistemic uncertainty (model disagreement): \newline
        $\sigma(x) = \sqrt{\frac{1}{T} \sum_{t=1}^T (\hat{y}_t(x) - \mu(x))^2}$ \\
        
        \raggedright\arraybackslash \textbf{Gaussian Process (GP)} & 
        The analytical posterior predictive mean, exacted via exact probabilistic conditioning on the training data. & 
        The analytical posterior predictive standard deviation, representing geometric uncertainty governed heavily by distance in the kernel space (e.g., RBF). \\
        
        \bottomrule
    \end{tabular}
    
    \vspace{1ex}
    \raggedright
    \footnotesize{\textit{Note:} For TabPFN, the $3.92$ scaling factor assumes the predicted quantiles approximate a normal distribution (where $2 \times 1.96 = 3.92$ covers 95\% of the density). For the Gaussian Process, both $\mu(x)$ and $\sigma(x)$ are derived analytically from the chosen covariance function.}
\end{table*}

\paragraph{TabPFN: A Foundation Model Surrogate forr ALMD:}
The Tabular Prior-Data Fitted Network (TabPFN) redefines tabular regression by framing it as a sequence-to-sequence problem, utilizing a Transformer architecture to implement In-Context Learning (ICL). Unlike traditional regressors that optimize parameters to fit a specific training set, TabPFN treats the entire dataset—consisting of feature vectors and their corresponding target values—as a sequence of input tokens. By prepending the labeled "support set" to an unlabeled "query" sample, the Transformer's self-attention mechanism computes the relational dependencies between the new material candidate and all previously observed experimental data points. This allows the model to perform a single forward pass that approximates the Posterior Predictive Distribution (PPD), effectively acting as a meta-learned Bayesian regressor.

TabPFN's inherent capability for probabilistic regression allows for robust uncertainty estimation, which, when coupled with its zero-training inference, makes it an ideal candidate for Active Learning in material design; the model can instantly suggest the next optimal experimental candidate as new data points are added to the context. Compared to traditional benchmarks, TabPFN offers the predictive power of gradient-boosted trees and the rigorous uncertainty quantification of Gaussian Processes, but with significantly higher efficiency and accuracy in the low-data regimes typical of experimental materials science. This architecture is particularly potent because it replaces the rigid structure of Random Forests or the kernel-dependency of Gaussian Processes with a flexible attention-based weighting, enabling it to capture complex, non-linear interactions even in the extreme low-data regimes typical of laboratory discovery.

\paragraph{Baseline surrogate model: Gaussian Process:} 

Gaussian Process (GP) has long been the standard for Active Learning (AL) based design \cite{vela2025gaussian,johnson2024active, alvi2024hierarchical, tian2025materials, alvi2025hierarchical} because it provides a mathematically rigorous Bayesian framework for uncertainty quantification. GP defines a prior over functions—typically assuming a specific kernel, such as the Radial Basis Function (RBF) or Matérn kernel, to encode smoothness and physical assumptions—and updates this prior into a posterior distribution as new experimental data is observed. This results in not only a mean prediction for a material's properties but also a formally derived variance that represents the model's confidence. In an active learning loop, this predictive variance is utilized by acquisition functions like Expected Improvement (EI) or Upper Confidence Bound (UCB) to navigate the exploration-exploitation tradeoff, systematically identifying material compositions that either maximize a target property or reduce the model's overall ignorance. While highly effective for small datasets, GP’s reliance on manually selected kernels and its $O(n^3)$ computational complexity can limit its flexibility in capturing complex, high-dimensional chemical relationships compared to modern attention-based models like TabPFN. We also found that GP has much lower regression performance compared to Random Forest and TabPFN.

\paragraph{Baseline surrogate model: Random Forest} 

Random Forest (RF) has been widely used in active learning material design \cite{johnson2024active, dale2025when, ma2025active, zhang2025active} due to its inherent robustness and its ability to provide heuristic uncertainty estimates through ensemble variance. In a typical AL material design loop, an RF model is trained on a sparse initial dataset of material compositions and properties; the "disagreement" or variance between the individual decision trees in the ensemble is then used as a proxy for epistemic uncertainty. This allows researchers to employ acquisition functions, such as Query-by-Committee or Upper Confidence Bound (UCB), to identify unlabeled material candidates in regions where the forest's predictions are least consistent. While widely adopted for its resistance to overfitting and its ability to handle high-dimensional chemical descriptors with minimal tuning, RF-based active learning often faces challenges in low-data regimes where the bootstrap samples may not provide a diverse enough "committee" for accurate uncertainty quantification—a limitation that more modern Bayesian approaches and foundation models like TabPFN aim to address.

\subsection{Benchmark Datasets }

We have evaluated the performance of AL algorithms based on different combination of surrogate models and acquisition functions over a set of dataset as shown in Table \ref{tab:datasets}.

The first three datasets concerns lattice thermal conductivity (LTC) prediction, comprising 3,148 crystal structures. Diamond (C) was explicitly excluded from the original dataset due to its exceptionally high LTC value, which would trivialize the active learning task by making it too easily identifiable as the optimal candidate. Three datasets were constructed using three different feature representations: elemental concentration (LTC\_C), Magpie features (LTC\_M), and a mixed representation combining Magpie with 20 structural features (LTC\_S).
All features were computed using the \textit{matminer} library \cite{ward2018matminer}. The resulting feature vector for each material is a concatenation of three distinct categories: (1) \textit{compositional features}, derived from the material's stoichiometry and incorporating Magpie elemental property statistics, average valence orbital configurations, and ionic property approximations; (2) \textit{structural features}, including bulk density metrics extracted directly from the crystallographic structure; and (3) \textit{symmetry features}, encoding the space group and rotational properties of the lattice. Any missing or non-numeric values encountered during featurization were imputed to zero to ensure numerical stability in downstream modeling.

The second group of datasets are on predicting hardness and and electrical conductivity of copper-based alloys \cite{gorsse2023dataset}. The hardness dataset (Cu\_HD\_C/M) has 1614 samples while the electrical conductivity dataset (Cu\_HD\_C/M) has 1826 samples. Both come with two different feature representations including element concentrations and Magpie features.

The third group of datasets are on predicting glass-forming ability (critical casting diameter (Dmax)) of Fe-based bulk metallic glasses \cite{bobadilla2025machine}. These are amorphous metal alloys, also known as metallic glasses that possess a unique combination of properties including very high elastic strain limit and excellent resistance to corrosion and wear, making them highly attractive for various industrial applications. The dataset has 495 samples and comes with three different representations including concentration (Glass\_DS1), physical features (Glass\_DS2), and Magpie features (Glass\_DS3).

\begin{table}[h]
\centering
\caption{Summary of 25 Benchmark Datasets}
\label{tab:datasets}
\begin{tabular}{llrrll}
\toprule
\textbf{Dataset Code } &\textbf{Dataset} & \textbf{Samples} & \textbf{Features} & Opt. Objective \\
\midrule
LTC\_C &LTC Concentration & 3,148 & 62 & LTC\\
LTC\_M &LTC Magpie & 3,148 & 126 & LTC\\
LTC\_S &LTC Structure & 3,148 & 146 & LTC\\

Cu\_HD\_C& Cu Alloy Hardness Concentration  & 1,614 & 34 & Hardness&\cite{gorsse2023dataset}\\
Cu\_HD\_M& Cu Alloy Hardness Magpie  & 1,614 & 132 & Hardness&\cite{gorsse2023dataset}\\
Cu\_EC\_C& Cu Alloy Elect. Conduct. Concentration  & 1,826 & 34 & Electrical Conductivity&\cite{gorsse2023dataset}\\
Cu\_EC\_M& Cu Alloy Elect. Conduct. Magpie  & 1,826 & 132 & Electrical Conductivity&\cite{gorsse2023dataset}\\

Glass\_DS1&Bulk glass forming alloy Concentration & 495 & 22 & glass forming &\cite{bobadilla2025machine}\\
Glass\_DS2&Bulk glass forming alloy  Physical & 495 & 14 & glass forming &\cite{bobadilla2025machine}\\
Glass\_DS3&Bulk glass forming alloy  Magpie & 495 & 145 & glass forming&\cite{bobadilla2025machine}\\

\bottomrule
\end{tabular}
\end{table}

We evaluate our active learning framework on ten molecular property prediction datasets derived from ChEMBL, spanning a diverse range of ADMET (absorption, distribution, metabolism, excretion, and toxicity) endpoints critical to drug discovery. The datasets vary substantially in size, ranging from 642 molecules (FreeSolv, hydration free energy) to 7,385 molecules (LD50, acute oral toxicity), and include: Caco-2 permeability (910), clearance microsome (1,102), clearance hepatocyte (1,213), VDSS volume of distribution (1,130), PPBR plasma protein binding rate (1,614), lipophilicity (4,200), LD50 (7,385), half-life (667), FreeSolv (642), and SARS-CoV-2 3CLpro inhibition (880). Each molecule is represented by 210 RDKit physicochemical descriptors, providing a unified feature space across all tasks. This collection covers a broad spectrum of biological and physicochemical properties — from membrane permeability and metabolic stability to antiviral activity and toxicity — making it a rigorous and diverse benchmark for evaluating active learning strategies in molecular machine learning.

\section{Results}
\label{sec:others}

\subsection{Performance evaluation of ICAL in active-learning based materials discovery}

We evaluate the performance of AL algorithms with three surrogate functions (GP, RF, TabPFN) combined with the six acquisition functions. For a given dataset, the pool-based materials discovery process starts with a given percentage of initial population of samples with the lowest property values and use the active learning cycles to locate the global optimum with batch size of 1. The extra evaluation steps during the AL cycles are used as the performance metric. For each surrogate model, we pick its best acquisition function based on the mean extra evaluations needed to locate the global optimum averaged across all 19 init levels (5\%–95\%). For GP and RF, we repeat the experiments for 20 times to count their randomness while TabPFN has no such randomness. We have tuned the hyper-parameters for the UCB and normalized UCB and normalized EI acquisition functions. We set the $\beta$ to 30 and $\xi$ to 0.2 across all experiments and all datasets for simplicity. 

\begin{figure}[ht]
  \centering
  \includegraphics[width=\linewidth]{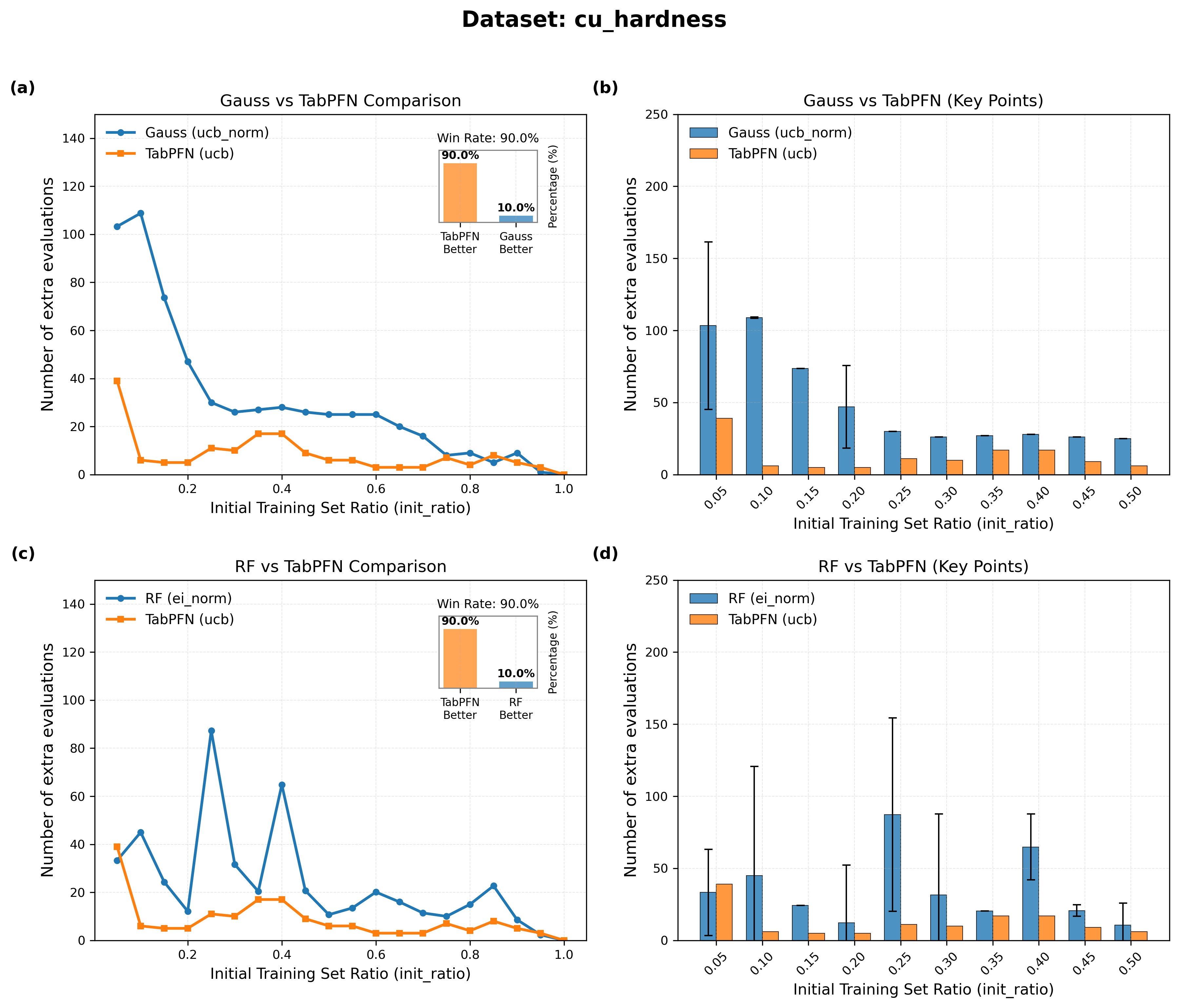}
  \caption{Performance of ICAL on Cu\_hardness dataset compared to GP and RF based AL. (a-b) Tabpfn against GP; (c-d): Tabpfn against RF.}
  \label{fig:cu_hardness}
\end{figure}

Figure~\ref{fig:cu_hardness} presents the active learning performance on the \textit{cu\_hardness} dataset, where the target property is copper alloy hardness and the metric is the number of extra evaluations required to identify the global optimum beyond the initial training set (lower is better). TabPFN with UCB acquisition achieves a 90\% win rate against both GP (UCB-norm) and RF (EI-norm) baselines --- the highest win rate observed across all datasets tested --- demonstrating a remarkably consistent and large performance advantage. As shown in panels (a) and (b), GP requires up to $\sim$110 extra evaluations at low init\_ratios (0.05--0.10), while TabPFN reduces this to $\sim$38 and $\sim$5 respectively, representing a 3--20$\times$ reduction in experimental effort in the most data-scarce regime. This advantage persists monotonically as init\_ratio increases, with TabPFN consistently requiring fewer than 20 extra evaluations across all conditions beyond init\_ratio $= 0.15$, while GP remains elevated at 25--50 evaluations through mid-range init\_ratios. The RF comparison (panels c and d) reveals an even more problematic failure mode: RF (EI-norm) exhibits severe instability, with multiple sharp spikes reaching $\sim$85 extra evaluations at init\_ratio $\approx 0.25$ and $\sim$65 at init\_ratio $\approx 0.40$. These spikes directly reflect RF's well-known weakness in small-data regimes, where its heuristic ensemble variance produces unreliable uncertainty estimates that mislead the EI acquisition function into querying uninformative or redundant compositions. Critically, panel (d) shows that RF's error bars are extraordinarily large --- extending to $\sim$155 at init\_ratio $= 0.25$ , meaning that in a single experimental campaign, a practitioner using RF could require over 150 additional synthesis-and-characterization cycles to find the optimal hardness alloy, at a potential cost exceeding \$300,000 at conservative per-sample estimates. In stark contrast, TabPFN maintains both low mean extra evaluations and tight error bars throughout, confirming that its meta-trained in-context posterior provides not only superior average efficiency but also reliable, reproducible convergence that is essential for practical materials discovery workflows.

\begin{figure}[ht]
      \begin{subfigure}[t]{0.5\textwidth}
        \includegraphics[width=\textwidth]{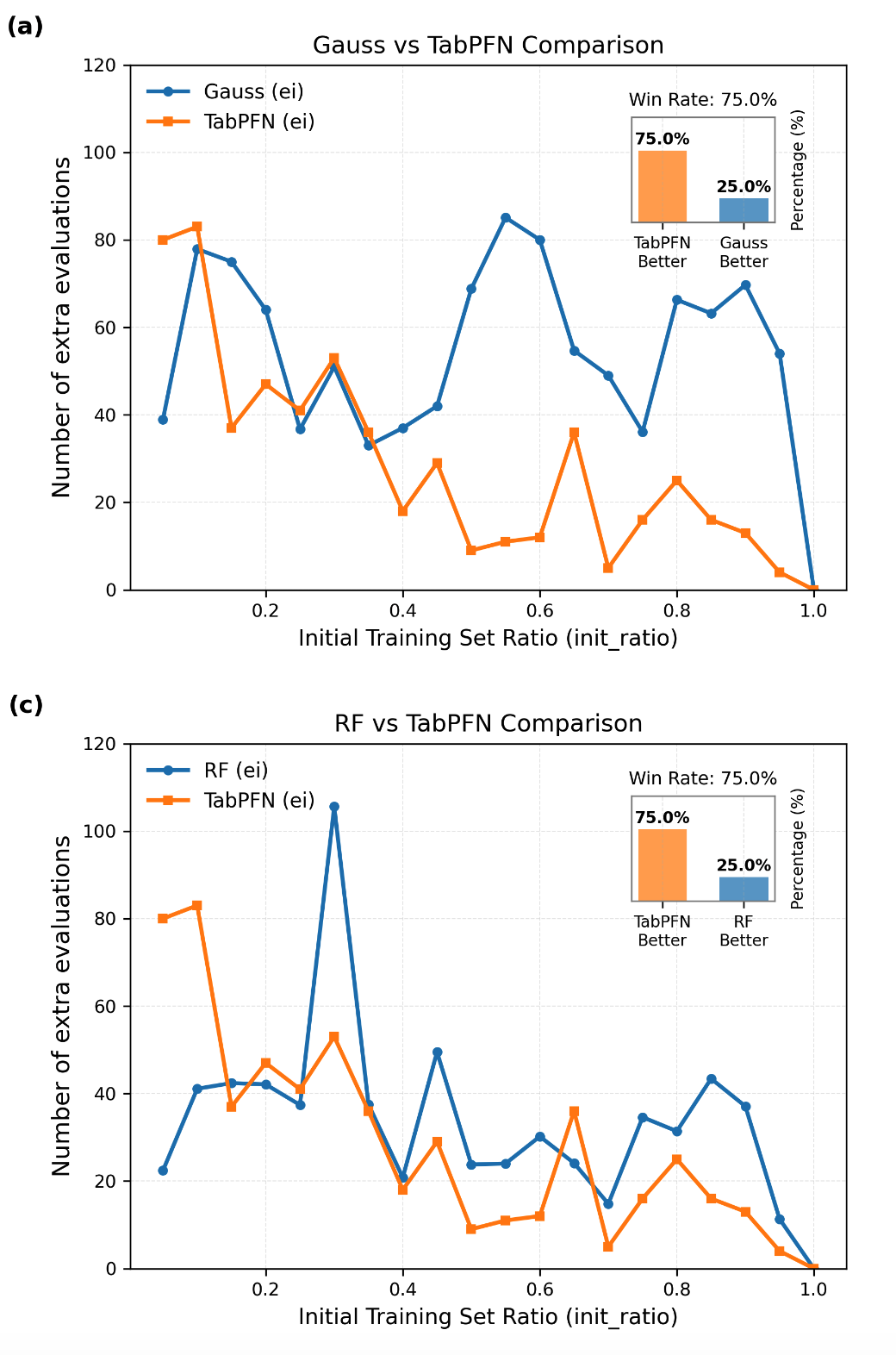}
        \vspace{-3pt}
        \label{fig:GaB3N4_predict2}
    \end{subfigure}
    \begin{subfigure}[t]{0.5\textwidth}
        \includegraphics[width=0.99\textwidth]{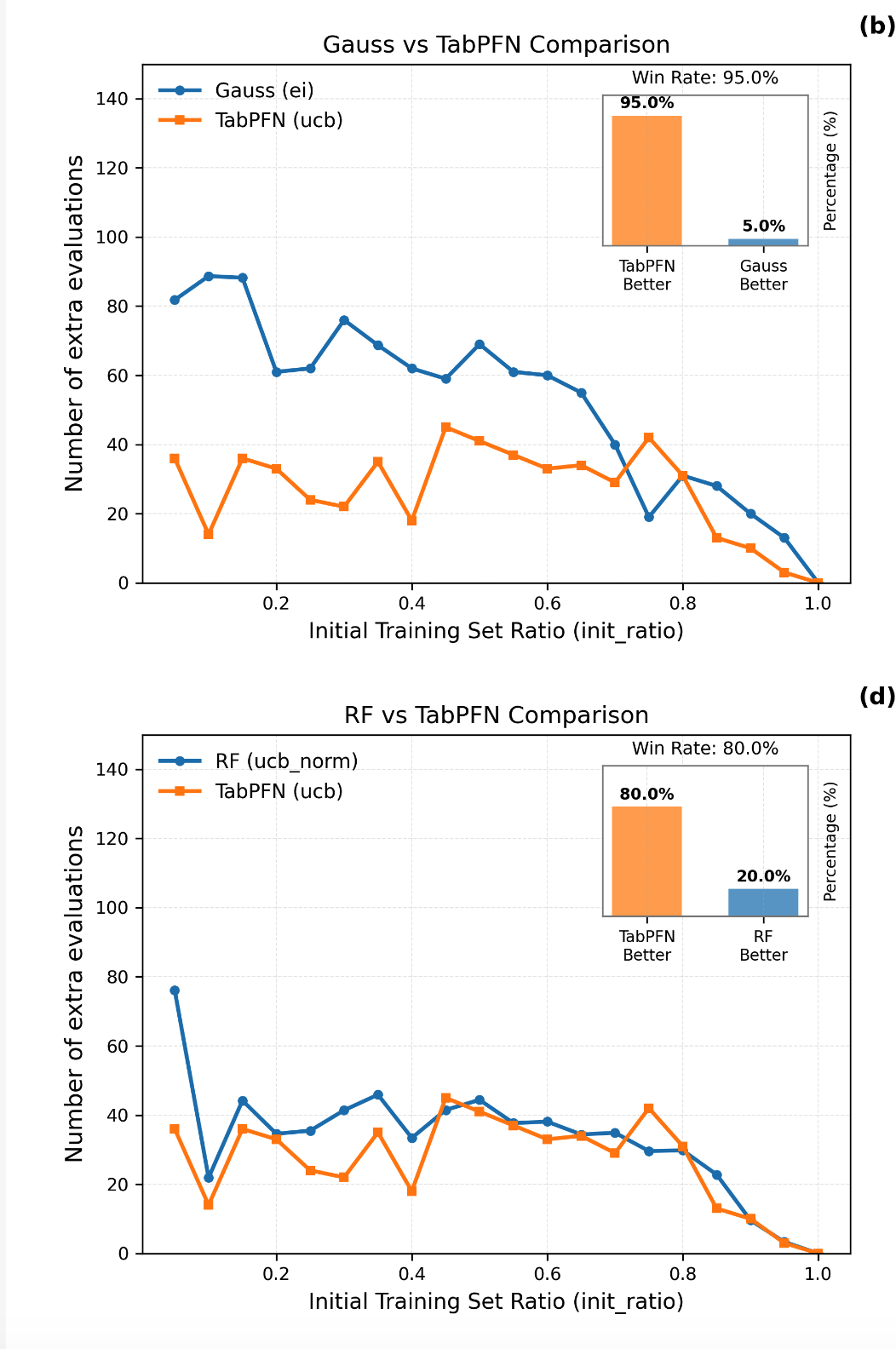}
        \vspace{-3pt}
        \label{fig:GaB2N3_predict1}
    \end{subfigure} 
  \caption{Performance of ICAL on (a) (c) Cu\_electric dataset and (b) (d) Glass\_DS3 dataset compared to GP and RF based AL.}
  \label{fig:cu_hardness_glass_magpie}
\end{figure}

Figure~\ref{fig:cu_hardness_glass_magpie} (a) shows the active learning results on the \textit{cu\_electric} dataset targeting electrical conductivity, where TabPFN (EI) achieves a \textbf{75\% win rate} against both GP (EI) and RF (EI) baselines. Unlike the cu\_hardness dataset where TabPFN's advantage was consistent across all init\_ratios, a notable exception emerges here at very low initial training set sizes (init\_ratio $\leq 0.15$): TabPFN requires $\sim$80--83 extra evaluations, slightly \textit{exceeding} GP's $\sim$38--79 and RF's $\sim$22--42 in this regime. This reversal suggests that TabPFN's in-context posterior requires a minimum amount of task-specific data to form a reliable representation of the conductivity landscape before its modeling advantage over GP and RF materializes. Beyond init\_ratio $\sim 0.20$, however, TabPFN's superiority becomes pronounced and sustained: GP remains persistently elevated at 40--85 extra evaluations with a notable peak at init\_ratio $\approx 0.55$ ($\sim$85), while TabPFN drops to below 15 evaluations across the same range. RF (panel c) exhibits its characteristic erratic behavior with a sharp spike to $\sim$105 at init\_ratio $\sim 0.30$, further confirming the unreliability of ensemble-variance-based uncertainty quantification in sparse data regimes. Both curves converge toward zero at init\_ratio $\to 1.0$, as expected when the training set covers nearly the entire candidate space. The crossover behavior at low init\_ratios is an important practical finding: it suggests that a minimum initial dataset size ,approximately 15--20\% of the candidate pool ,should be collected before deploying TabPFN-based active learning for electrical conductivity optimization.

Figure~\ref{fig:cu_hardness_glass_magpie} (b) (d) presents results on the \textit{glass\_DS3} dataset with magpie features targeting glass forming capability of alloys. TabPFN (UCB) achieves its highest win rate across all datasets tested: \textbf{95\%} against GP (EI) and \textbf{80\%} against RF (UCB-norm), suggesting that the glass forming capability landscape is particularly well-suited to TabPFN's in-context Bayesian inference. Identifying glass-forming alloys is notoriously challenging due to the complex, nonlinear interplay between compositional features and the tendency to avoid crystallization --- precisely the kind of structure that GP's rigid kernel assumptions fail to capture. As shown in the upper panel, GP requires 60--90 extra evaluations across nearly the entire init\_ratio range of 0.05--0.65, while TabPFN stays consistently below 45 throughout, representing a \textbf{2--4$\times$ reduction} in the number of alloy synthesis and characterization experiments needed. This is practically significant given that producing and validating a glass-forming alloy ( involving arc melting, melt spinning, and structural characterization via XRD to confirm amorphous phase formation )is among the more expensive experimental workflows in alloy design. The RF comparison (lower panel) tells a more nuanced story: RF and TabPFN perform comparably in the mid-range (init\_ratio $= 0.45$--$0.75$), with curves frequently crossing, which accounts for RF's 20\% win rate --- the highest any baseline achieves across all datasets. Nevertheless, TabPFN maintains a decisive advantage at low init\_ratios where it matters most, with RF spiking to $\sim$77 extra evaluations at init\_ratio $= 0.05$ while TabPFN requires only $\sim$37, and TabPFN's overall smoother trajectory confirming more reliable convergence toward the optimal glass-forming composition.

\begin{figure}[htb]
  \centering
  \includegraphics[width=\linewidth]{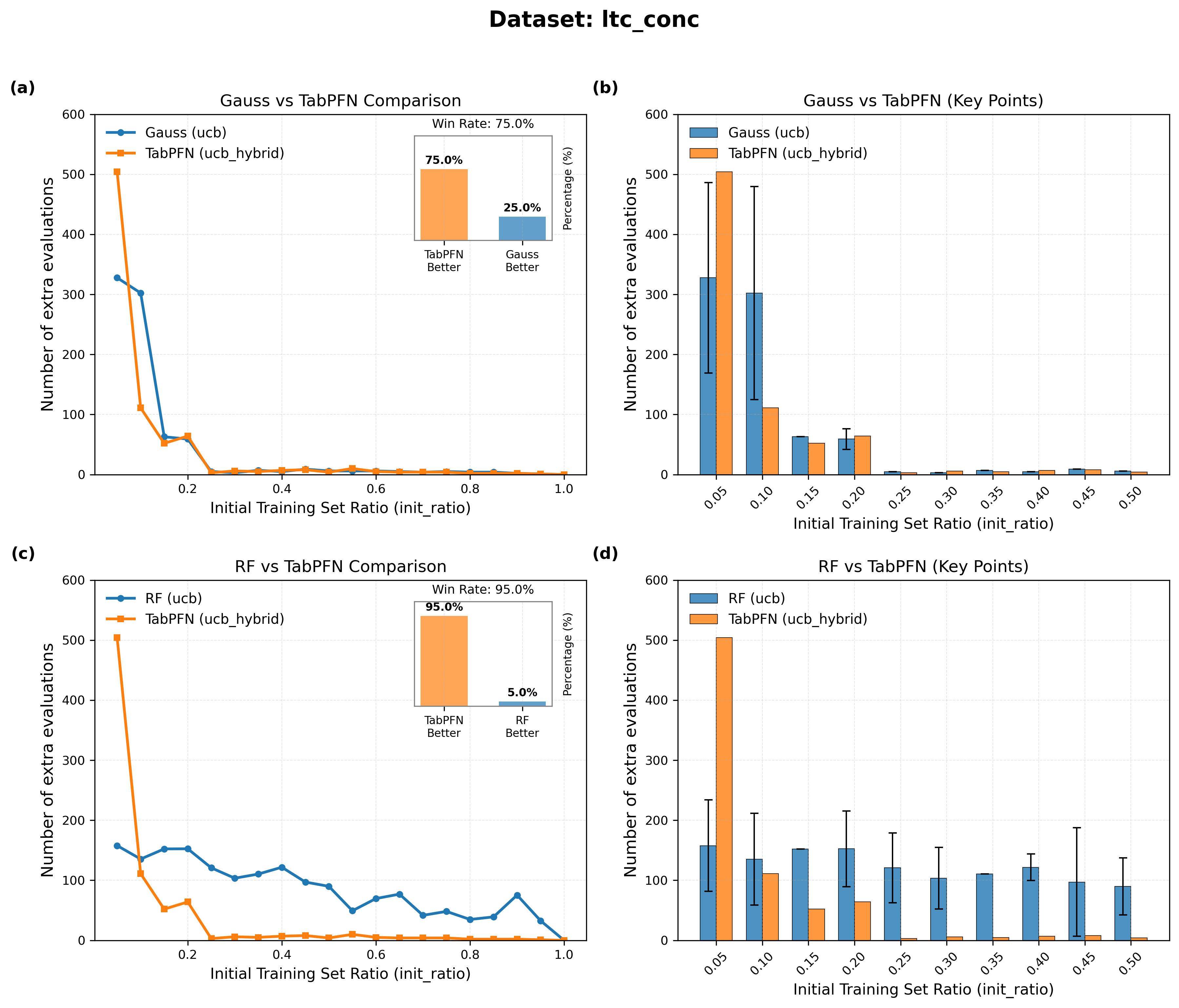}
  \caption{Performance of ICAL on LTC\_conc dataset with element concentration features compared to GP and RF based AL.}
  \label{fig:ltc_conc}
\end{figure}

Figure~\ref{fig:ltc_conc} presents active learning results on the \textit{ltc\_conc} dataset targeting the lattice thermal conductivity (LTC) of crystal materials described by concentration features. Several distinctive and noteworthy patterns emerge that set this dataset apart from the others. TabPFN (UCB\_hybrid) achieves a 75\% win rate against GP (UCB) and a remarkable 95\% win rate against RF (UCB) (the highest RF win rate across all datasets tested),indicating that RF is particularly ill-suited for navigating the LTC--composition landscape of crystal materials when only concentration features are used.
The most striking observation in this dataset is the dramatic collapse to fewer than 10 extra evaluations beyond init\_ratio $\approx 0.20$: both GP and TabPFN achieve this near-perfect convergence all the way to init\_ratio $= 1.0$ (panels a and b). This suggests that the LTC landscape of crystal materials parameterized by concentration features is relatively low-complexity and well-structured --- once roughly 20\% of the compositional space is sampled, the global optimum becomes straightforward to locate for both GP and Tabpfn surrogates. This is physically intuitive: LTC in crystals is governed primarily by phonon scattering mechanisms, which are strongly correlated with atomic mass contrast and bond stiffness (both of which vary smoothly with elemental concentration ), producing a relatively smooth and navigable property landscape compared to mechanical properties or glass-forming ability. However, this result is still surprising as accurate prediction of LTC is extremely difficult by itself, showing the different ML model capability required for AL.

However, the critical distinction between TabPFN and GP lies in the extremely data-scarce regime (init\_ratio $= 0.05$--$0.10$), where the difference is both large and practically consequential. At init\_ratio $= 0.05$, GP requires $\sim$330 extra evaluations while TabPFN requires $\sim$500 --- a rare case where TabPFN is \textit{worse} at the smallest init size, consistent with the minimum data threshold effect observed in the cu\_electric dataset. At init\_ratio $= 0.10$, however, the roles reverse sharply: GP still needs $\sim$305 extra evaluations while TabPFN drops precipitously to $\sim$110, a nearly 3$\times$ reduction. This crossover between init\_ratio $= 0.05$ and $0.10$ defines a critical threshold: below it, TabPFN lacks sufficient in-context examples to infer the LTC--composition mapping of the crystal system; above it, its superior modeling capacity enables rapid identification of the optimum.

The RF comparison (panels c and d) is even more revealing. RF (UCB) fails catastrophically and persistently: it requires 100--160 extra evaluations across the entire range from init\_ratio $= 0.15$ to $0.50$ (panel d), never converging to fewer than 10 extra evaluations even when 50\% of the candidate crystal compositions have been labeled. This is a qualitatively different failure mode from GP --- RF does not merely need more data to converge, it fails to converge efficiently at all in this crystal composition feature space. The large error bars on RF in panel (d) further confirm that RF's UCB-guided search is highly sensitive to random seed, producing wildly inconsistent outcomes across runs which is a serious liability when synthesizing and characterizing crystal materials carries significant cost and effort. TabPFN, by contrast, converges to fewer than 10 extra evaluations by init\_ratio $= 0.25$ with tight variance, demonstrating that the UCB\_hybrid acquisition combined with TabPFN's calibrated uncertainty is both efficient and reproducible for crystal LTC optimization.

\paragraph{Importance of representation}

Due to the extremely small datasets in typical materials AL, we find that the AL performancce strongly depends on the features used in surrogate models. To investigate this issue, we test different features for the four datasets cu\_electric, cu\_hardness, glass forming alloy, and LTC. We calculate the mean number of extra evaluations per surrogate model with its best acquisition function.

\begin{table}[htbp]
\centering
\caption{Mean number of extra evaluations per surrogate model averaged across all 19 init\_ratio levels (5\%--95\%), using each model's best acquisition function for that dataset. Percentage of extra evaluation saving achieved by TabPFN relative to GP and RF are shown in the final two columns; negative values indicate TabPFN underperforms the baseline (only for the inappropriate magppie feature sets).}
\label{tab:summary}
\begin{tabular}{lcccclcc}
\toprule
\textbf{Dataset} & \textbf{Gauss GP} & \textbf{RF} & \textbf{TabPFN} & \textbf{Winner} & \makecell{\textbf{Best Acq} \\ \textbf{(TabPFN)}} & \makecell{\textbf{\% Saving.} \\ \textbf{(vs GP)}}  & \makecell{\textbf{\% Saving.} \\ \textbf{(vs RF)}}  \\
\midrule
cu\_electric         & 58.5            & 37.5           & {[}\textbf{30.1}{]} & TabPFN & ei          & \textbf{48.5\%}    & \textbf{19.7\%}   \\
cu\_electric\_magpie & \textbf{82.5}   & 90.3           & 83.9                & Gauss  & ucb\_norm   & $-$1.7\%  & 7.1\%    \\
\midrule
cu\_hardness         & 31.3            & 18.8           & {[}\textbf{8.8}{]}  & TabPFN & ucb         & \textbf{71.9\%}    & \textbf{53.2\%}   \\
cu\_hardness\_magpie & \textbf{21.9}   & 23.5           & 24.1                & Gauss  & ei\_hybrid  & $-$10.0\% & $-$2.6\% \\
\midrule
glass\_ds1\_conc     & 72.4            & 66.0           & {}\textbf{48.8}{} & TabPFN & ucb\_norm   & 32.6\%    & 26.1\%   \\
glass\_ds2\_physical & 37.6            & 43.7           & {}\textbf{32.6}{} & TabPFN & ucb\_norm   & 13.3\%    & 25.4\%   \\
glass\_ds3\_magpie   & 67.5            & 31.6           & {[}\textbf{28.2}{]} & TabPFN & ucb         & \textbf{58.2\%}    & \textbf{10.8\%}   \\
\midrule
ltc\_conc            & 56.55           & 61.8           & {[}\textbf{39.9}{]} & TabPFN & ucb\_hybrid & \textbf{29.4\%}    & \textbf{35.4\%}   \\
ltc\_magpie          & 243.88          & 44.17          & \textbf{41.70}      & TabPFN & ucb         & 82.9\%    & 5.6\%    \\
ltc\_structure       & 237.37          & \textbf{55.05} & 84.1                & RF     & ei          & 64.6\% & $-$52.8\%\\
\midrule
\makecell{{Mean saving of} \\ {Extra Evaluations by TabPFN}}  &                 &  &                 &      &           & 52\% & 29.77\%\\

\bottomrule
\end{tabular}
\end{table}

Table~\ref{tab:summary} summarizes the mean number of extra evaluations per model averaged across all 19 init\_ratio levels (5\%--95\%), using each model's best acquisition function per dataset. TabPFN achieves the lowest extra evaluations on 8 out of 10 datasets, establishing it as the consistently dominant surrogate across diverse material classes and property targets. Averaged across all TabPFN-winning datasets, TabPFN reduces extra evaluations by 35--83\% relative to GP and 10--53\% relative to RF, representing substantial savings in experimental cost.

A particularly important finding emerges when comparing concentration-based versus Magpie feature sets for the copper and LTC datasets. For cu\_electric, TabPFN with concentration features requires only 30.1 extra evaluations on average, compared to 58.5 (GP) and 37.5 (RF) --- reductions of 49\% and 20\% respectively. However, switching to the high-dimensional Magpie feature set (cu\_magpie\_electric, $>$120 features) reverses the winner entirely: GP becomes best at 82.5 extra evaluations, while TabPFN degrades to 83.9 --- nearly indistinguishable from GP and worse than both concentration-feature models. A similar pattern holds for cu\_hardness: TabPFN achieves only 8.8 extra evaluations with concentration features (reductions of 72\% vs.\ GP and 53\% vs.\ RF), but with Magpie features (cu\_magpie\_hardness) GP wins at 21.9, and all three models perform substantially worse than their concentration-feature counterparts. The same trend is observed for LTC crystal materials: ltc\_conc gives TabPFN a win at 39.9 extra evaluations (reductions of 29\% vs.\ GP and 35\% vs.\ RF), whereas ltc\_magpie produces a dramatically inflated GP count of 243.88 (a 6$\times$ degradation relative to ltc\_conc )before TabPFN partially recovers to 41.70. These results consistently indicate that high-dimensional Magpie features introduce severe overfitting in GP's kernel estimation and destabilize RF's ensemble variance, undermining the uncertainty quantification that drives acquisition function decisions. TabPFN is more robust to this feature dimensionality increase, but still performs best with the leaner concentration feature representation. This additionally shows that the AL needs different model capability than the regular regression performance to achieve high performance: the magpie feature can obtain better LTC prediction models, but not necessarily better AL performance.

In contrast, the glass-forming alloy datasets reveal a strikingly different trend with respect to feature complexity. All three glass datasets are won by TabPFN, with mean extra evaluations of 48.8 (glass\_ds1, concentration features), 32.6 (glass\_ds2, physical features), and 28.2 (glass\_ds3, Magpie features) respectively. Notably, performance \textit{improves} monotonically as feature richness increases --- the opposite of what is observed for copper and LTC datasets. The Magpie feature set yields the best result for glass-forming ability prediction, reducing extra evaluations by 58\% relative to GP (67.5) and 11\% relative to RF (31.6). This suggests that glass-forming ability is governed by a more complex, nonlinear composition--property relationship that genuinely benefits from richer physicochemical descriptors: predicting whether an alloy will form an amorphous phase depends on subtle electronic structure, atomic size mismatch, and thermodynamic factors that concentration features alone cannot adequately encode, but which Magpie descriptors partially capture. This stands in direct contrast to copper hardness and LTC, where the property landscape is smoother and high-dimensional Magpie features introduce more noise than signal. The feature-set sensitivity thus appears to be material-class dependent, and the consistent superiority of TabPFN across all three glass datasets ( regardless of feature type ) further demonstrates its robustness to feature space variations. Finally, ltc\_structure is the only dataset where RF wins (84.1 extra evaluations for TabPFN vs.\ 55.05 for RF), suggesting that structural features present a unique inductive bias that RF's decision-tree architecture exploits more effectively than TabPFN's tabular meta-prior in this specific setting.

We also calculate the percentages of saving by using TabPFN based AL. Across the 10 datasets evaluated, TabPFN achieves a mean saving of 52\% in extra evaluations relative to GP and 29.77\% relative to RF when averaged over the datasets where TabPFN wins with the best-performing feature set, demonstrating its consistent superiority as a surrogate model for active learning-based materials discovery.

For the copper alloy datasets, the choice of feature set critically determines performance. With concentration features, TabPFN reduces extra evaluations by 48.5\% vs.\ GP and 19.7\% vs.\ RF for electrical conductivity (cu\_electric), and by an even larger margin of 71.9\% vs.\ GP and 53.2\% vs.\ RF for hardness (cu\_hardness) --- the largest reduction against RF across all datasets. These results confirm that concentration features provide the best-performing representation for copper alloy properties, and the corresponding Magpie feature variants (cu\_electric\_magpie, cu\_hardness\_magpie) are excluded from the mean saving calculation as they do not represent the optimal feature set for most surrogate models on these material systems.

For the glass-forming alloy datasets, TabPFN wins across all three feature representations, but performance improves monotonically with feature richness. The Magpie feature set (glass\_ds3\_magpie) yields the best TabPFN result with 28.2 extra evaluations, achieving savings of 58.2\% vs.\ GP and 10.8\% vs.\ RF, and is therefore selected as the best-performing feature set for this material class. The physical feature set (glass\_ds2\_physical) offers a balanced alternative with savings of 13.3\% vs.\ GP and 25.4\% vs.\ RF, while the concentration feature set (glass\_ds1\_conc) provides the most modest gains at 32.6\% vs.\ GP and 26.1\% vs.\ RF. The progressive improvement from concentration to physical to Magpie features suggests that glass-forming ability is governed by complex physicochemical interactions that benefit from richer descriptors, in contrast to copper alloy properties.

For the lattice thermal conductivity (LTC) datasets, TabPFN with concentration features (ltc\_conc) achieves savings of 29.4\% vs.\ GP and 35.4\% vs.\ RF and is selected as the best-performing feature set for the mean saving calculation, as the Magpie variant (ltc\_magpie) offers only marginal additional improvement over RF (5.6\%) despite a dramatic 82.9\% reduction vs.\ GP ,the latter being inflated by GP's catastrophic failure with high-dimensional Magpie features rather than a genuine improvement in the active learning loop. The ltc\_structure dataset is the sole case where TabPFN underperforms, with RF winning at 55.05 extra evaluations vs.\ TabPFN's 84.1, suggesting that structural features encode an inductive bias that RF's decision-tree architecture exploits more effectively than TabPFN's tabular meta-prior; this dataset is excluded from the mean saving calculation accordingly. Taken together, the feature set analysis reveals a consistent principle: TabPFN performs best with the leanest feature representation that adequately captures the relevant physics of the target property, while high-dimensional feature sets only benefit TabPFN when the property landscape is sufficiently complex to warrant richer descriptors, as in the case of glass-forming ability.

\subsection{Mechanism Analysis: Explaining TabPFN's active-learning performance}

The efficiency of an active learning framework is fundamentally limited by the surrogate model's ability to balance exploitation (via accurate mean prediction) and exploration (via calibrated uncertainty estimates). 
To evaluate the underlying mechanisms driving the active learning performance, we conducted a comprehensive assessment of the regression accuracy and uncertainty quantification (UQ) capabilities of the three surrogate models (TabPFN, RF, and GP) over the Cu-alloy electric conductivity dataset, as summarized in Table~\ref{tab:uq_performance}. All results are based on 5-fold cross-validation. Beyond standard error metrics like RMSE and $R^2$, we specifically prioritized the Spearman Rank Correlation ($\rho$) and Negative Log-Likelihood (NLL). In the context of materials screening, the Spearman metric is particularly critical because active learning acquisition functions (e.g., Expected Improvement) rely on the \textit{relative ranking} of candidates to prioritize the next set of experiments, rather than their absolute predicted values.

\begin{table}[ht]
    \centering
    \caption{Uncertainty Quantification \& Regression Performance Comparison. Values are reported as Mean $\pm$ Standard Deviation. Arrows indicate whether lower ($\downarrow$) or higher ($\uparrow$) values are better. All are 5-Fold Cross-Validation Results for Cu-alloy Electrical Conductivity dataset (\%IACS). Best results are highlighted in \textbf{bold}.}
    \label{tab:uq_performance}
    \resizebox{\textwidth}{!}{%
    \begin{tabular}{lrrrrrrr}
        \toprule
        \textbf{Model} & \textbf{RMSE} $\downarrow$ & \textbf{$R^2$} $\uparrow$ & \textbf{Spearman} $\uparrow$ & \textbf{NLL} $\downarrow$ & \textbf{PICP} (0.95) & \textbf{MPIW} $\downarrow$ & \textbf{AUSE} $\downarrow$ \\
        \midrule
        Gaussian Process & $8.38 \pm 0.51$ & $0.73 \pm 0.04$ & $0.81 \pm 0.02$ & $3.46 \pm 0.03$ & $\mathbf{0.93 \pm 0.01}$ & $32.50 \pm 1.02$ & $5.58 \pm 0.16$ \\
        Random Forest    & $5.86 \pm 0.28$ & $0.87 \pm 0.02$ & $0.89 \pm 0.02$ & $3.11 \pm 0.25$ & $0.87 \pm 0.02$ & $15.93 \pm 0.35$ & $1.77 \pm 0.42$ \\
        TabPFN           & $\mathbf{5.56 \pm 0.42}$ & $\mathbf{0.88 \pm 0.03}$ & $\mathbf{0.92 \pm 0.01}$ & $\mathbf{2.34 \pm 0.15}$ & $0.88 \pm 0.02$ & $\mathbf{11.16 \pm 0.34}$ & $\mathbf{0.83 \pm 0.12}$ \\
        \bottomrule
    \end{tabular}
    }
\end{table}

 As shown in the Table \ref{tab:uq_performance}, TabPFN achieves a dominant regression performance with the lowest RMSE (5.56) and the highest Spearman correlation (0.92), significantly outperforming the Gaussian Process ($\rho=0.81$). This indicates that TabPFN serves as a superior ``exploitation'' engine, ensuring that the candidates identified as high-value are statistically more likely to be true positives compared to the baselines.

In terms of uncertainty quantification, which governs the ``exploration'' efficiency of the active learning loop, we utilized the NLL, Prediction Interval Coverage Probability (PICP), Mean Prediction Interval Width (MPIW), and the Area Under the Sparsification Error (AUSE) curve. PICP measures the probability that the true value falls within the predicted confidence interval (target 0.95), while MPIW measures the sharpness of those intervals. A useful uncertainty estimate must balance high coverage with narrow width. 
First we found that TabPFN achieved the lowest NLL (2.34) compared to the Gaussian Process (3.46) and Random Forest (3.11), a result that is significant because NLL is a proper scoring rule that penalizes both prediction error and poor uncertainty calibration; this confirms that TabPFN's predicted posterior distributions are the most faithful to the true data distribution. Next, while the Gaussian Process achieved high coverage (PICP $\approx$ 0.93), it did so by producing excessively wide intervals (MPIW $\approx$ 32.50), rendering the uncertainty signal uninformative for distinguishing between candidates. In contrast, TabPFN achieved comparable coverage with 3x sharper intervals (MPIW $\approx$ 11.16) and the lowest AUSE (0.83) compared to 15.58 of GP and 1.77 of RF. The low AUSE score is particularly telling, as it quantifies the correlation between estimated uncertainty and actual prediction error; a low score implies that the model is ``honestly'' uncertain only when it is likely to be wrong. This superior calibration allows the TabPFN-driven acquisition function to efficiently navigate the chemical space, avoiding the ``blind'' exploration often caused by the over-conservative uncertainty estimates of standard Gaussian Processes.

We also evaluated the peformances of TabPFN and the baseline ML models over the more challenging material design for higher LTC using the concentration features as shown in Table \ref{tab:cv_ltc}, which presents five-fold cross-validation results evaluating both predictive accuracy and uncertainty quantification quality for the three surrogate models on the LTC dataset with concentration features. While all three models achieve comparable predictive accuracy (TabPFN and RF both attain RMSE $\approx$ 48.7 and R$^2$ = 0.18, modestly outperforming GP whose R$^2$ of $-$0.01 indicates near-zero predictive power), the critical differentiator lies in the quality of uncertainty estimation, which directly governs active learning query selection. GP's near-perfect PICP of 0.99 is deceptive: it is achieved trivially through extremely wide prediction intervals (MPIW = 237.23, nearly 10$\times$ larger than TabPFN's 33.16), meaning GP's uncertainty covers almost the entire output range regardless of input and provides no meaningful signal for distinguishing informative from uninformative candidates. This pathological overestimation of uncertainty explains GP's poor active learning performance on ltc\_conc in Table~\ref{tab:summary}, where it requires 56.55 extra evaluations on average --- 29.4\% more than TabPFN. RF achieves a tighter MPIW of 24.37 but significantly undercoveres at PICP = 0.86, falling well short of the nominal 95\% level and indicating that its heuristic ensemble variance systematically underestimates true predictive uncertainty, which in turn causes the acquisition function to over-exploit overconfident predictions and miss the global optimum , consistenting with RF requiring 61.8 extra evaluations in Table~\ref{tab:summary}. TabPFN achieves the best balance: a PICP of 0.94 closest to the nominal 95\% coverage, a well-controlled MPIW of 33.16, the lowest NLL of 3.22 (less than half of GP's 5.81 and less than half of RF's 7.53), and the lowest AUSE of 2.16 --- less than half of RF's 4.50 and five times lower than GP's 10.80. The AUSE metric is particularly relevant here as it directly measures how well-ordered the uncertainty estimates are for active learning: a low AUSE means that samples flagged as uncertain by the surrogate are indeed the ones with the largest prediction errors, enabling the acquisition function to query the most informative compositions efficiently. This superior uncertainty calibration is the mechanistic explanation for TabPFN's 29.4\% and 35.4\% savings in extra evaluations relative to GP and RF respectively on ltc\_conc in Table~\ref{tab:summary}, and underscores that predictive accuracy alone is insufficient to characterize surrogate quality for active learning because reliable and calibrated uncertainty quantification is equally, if not more, important.

\begin{table}[htbp]
\centering
\caption{Five-fold cross-validation results on the LTC dataset with concentration features. 
Values are reported as mean $\pm$ standard deviation across folds. Arrows indicate whether 
lower ($\downarrow$) or higher ($\uparrow$) values are better. PICP: prediction interval 
coverage probability; MPIW: mean prediction interval width; AUSE: area under the 
sparsification error curve. Best results are highlighted in bold.}
\label{tab:cv_ltc}
\small
\begin{tabular}{lccccccc}
\toprule
\textbf{Model} & \textbf{RMSE $\downarrow$} & \textbf{R$^2$ $\uparrow$} & \textbf{Spearman $\uparrow$} & \textbf{NLL $\downarrow$} & \textbf{PICP (95\%)} & \textbf{MPIW $\downarrow$} & \textbf{AUSE $\downarrow$} \\
\midrule
Gaussian Process & 53.26 $\pm$ 33.75 & $-$0.01 $\pm$ 0.01          & 0.61 $\pm$ 0.04          & 5.81 $\pm$ 1.08          & \textbf{0.99 $\pm$ 0.00} & 237.23 $\pm$ 39.87          & 10.80 $\pm$ 3.56          \\
Random Forest    & 48.62 $\pm$ 32.22 & 0.18 $\pm$ 0.06              & 0.83 $\pm$ 0.01          & 7.53 $\pm$ 3.42          & 0.86 $\pm$ 0.02          & \textbf{24.37 $\pm$ 2.74}   & 4.50 $\pm$ 1.84           \\
TabPFN           & \textbf{48.86 $\pm$ 32.78} & \textbf{0.18 $\pm$ 0.06} & \textbf{0.86 $\pm$ 0.01} & \textbf{3.22 $\pm$ 0.32} & 0.94 $\pm$ 0.01          & 33.16 $\pm$ 3.47            & \textbf{2.16 $\pm$ 0.43}  \\
\bottomrule
\end{tabular}
\end{table}

\FloatBarrier

\section{Discussion and Conclusion}

In this work, we proposed and evaluated the In-Context Active Learning (ICAL) algorithm, which 
leverages TabPFN --- a transformer-based foundation model pre-trained to approximate Bayesian 
inference over tabular data --- as a principled surrogate for pool-based active learning in 
materials discovery. Benchmarked across 10 datasets spanning copper alloys, bulk metallic glasses, 
and crystal lattice thermal conductivity, ICAL achieves a mean saving of 52\% in extra evaluations 
relative to GP and 29.77\% relative to RF, winning on 8 out of 10 datasets. These results 
establish TabPFN as a consistently superior surrogate that resolves the fundamental tension between 
modeling capacity and uncertainty reliability that has constrained GP and RF-based active learning 
in materials science.

These results suggest that foundation models can address the small-data surrogate modeling challenge in experimental discovery.

Our mechanistic analysis reveals that TabPFN's active learning advantage is not merely a consequence 
of better regression accuracy --- indeed, TabPFN and RF achieve comparable RMSE and R$^2$ on 
several datasets --- but rather stems from its superior uncertainty quantification. By meta-learning 
a universal prior over tabular regression tasks from millions of synthetic datasets, TabPFN produces 
well-calibrated predictive posteriors that GP and RF, both trained from scratch on scarce 
experimental data, cannot replicate. This is particularly consequential in the early iterations of 
an AL campaign, where the training set is smallest and the quality of uncertainty estimates most 
directly determines whether the acquisition function explores productively or wastes evaluations on 
uninformative candidates. The consistently lower NLL and AUSE scores achieved by TabPFN across 
both the Cu-alloy and LTC datasets confirm that its in-context posterior faithfully represents 
predictive uncertainty in precisely these critical low-data regimes.

Several important practical insights emerge from our experiments. First, feature representation 
is a decisive and often underappreciated factor in AL performance, independent of regression 
accuracy: high-dimensional Magpie features degrade AL performance for copper alloys and LTC 
despite improving standalone regression metrics, while the same features benefit glass-forming 
alloy discovery where the property landscape is more complex. This finding suggests that feature 
selection for AL should be guided by AL performance metrics rather than cross-validation accuracy 
alone. Second, a minimum initial dataset size of approximately 10--20\% of the candidate pool 
is required before TabPFN's in-context advantage fully materializes, below which insufficient 
context examples limit its ability to infer the composition--property mapping. Practitioners 
should account for this threshold when designing AL campaigns with very small initial budgets. 
Third, our proposed normalized and hybrid acquisition functions effectively address the scale 
instability that plagues standard EI and UCB in small-data regimes, and are broadly applicable 
regardless of surrogate model choice.

The broader implications of ICAL extend well beyond the alloy and crystal datasets evaluated 
here. TabPFN's pre-trained small-data regression and uncertainty quantification capabilities 
make ICAL a natural fit for any AL application where labeled data is expensive to acquire and 
datasets are small. In molecule discovery and drug design, where the cost of synthesizing and 
assaying candidate compounds is high and datasets are routinely small \cite{Tang2024, Marin2024, 
tosh2025bayesian}, ICAL could accelerate the identification of lead compounds by reducing the 
number of wet-lab evaluations required to locate optimal candidates. The regression-based active 
learning framework we adopt is directly analogous to virtual screening workflows used in 
ultra-large library docking \cite{Marin2024}, and TabPFN's single-pass inference makes it 
computationally practical even when acquisition decisions must be made rapidly across millions 
of candidates. In catalyst design, where active learning has already demonstrated order-of-magnitude 
reductions in experimental effort \cite{suvarna2024active, moon2024active, rohr2020benchmarking}, 
ICAL's improved uncertainty calibration could further sharpen the exploration--exploitation 
balance and reduce the number of costly electrochemical characterization experiments. 

ICAL is also well-positioned for integration into self-driving laboratory (SDL) platforms 
\cite{tom2024self, canty2025science, szymanski2023autonomous, Abolhasani2023}, where the 
surrogate model must make reliable query decisions in real time with continuously growing but 
always small labeled datasets. TabPFN's zero-retraining inference can perform a complete 
Bayesian update in a single forward pass as new experimental results are added to the context, which aligns naturally with the closed-loop, iterative experimentation paradigm of SDLs, 
eliminating the retraining latency that can become a bottleneck in high-throughput autonomous 
workflows. Furthermore, ICAL can be straightforwardly extended from pool-based screening to 
open-ended generative materials design \cite{chen2026survey, lookman2026materials}, where 
candidate materials are proposed by a generative model and TabPFN serves as the screening 
surrogate to prioritize which generated candidates are worth evaluating experimentally.

Several limitations of the current work point toward promising directions for future research. 
The performance reversal on the \textit{ltc\_structure} dataset, where RF outperforms TabPFN, 
suggests that structural features encode an inductive bias that TabPFN's tabular meta-prior 
does not currently capture. Extending TabPFN's pre-training distribution to include 
structure-aware representations, or developing hybrid architectures that combine TabPFN's 
probabilistic inference with graph neural network encoders for structural features, could 
address this limitation. The identified minimum data threshold effect also motivates the 
development of adaptive initialization strategies that ensure sufficient compositional coverage 
before the AL loop begins. Finally, while we evaluate single-objective AL in this work, 
extending ICAL to multi-objective settings \cite{ji2024multi} represents a natural and practically important 
extension that we leave for future work, where practitioners 
simultaneously optimize multiple competing properties such as strength and ductility, or 
conductivity and corrosion resistance.

In conclusion, ICAL establishes TabPFN-based in-context Bayesian active learning as a 
principled, practical, and broadly applicable framework for data-efficient materials discovery. 
By resolving the long-standing tension between modeling capacity and uncertainty reliability 
in small-data surrogate modeling, ICAL reduces the number of expensive experiments needed 
to find optimal materials by up to 72\% relative to GP and 53\% relative to RF on individual 
datasets, with mean savings of 52\% and 29.77\% respectively across all benchmarks. Given 
the \$500--\$20{,}000 cost per synthesis-and-characterization experiment in alloy and crystal 
materials research, these savings represent not only a methodological advance but a tangible 
reduction in the financial and human cost of materials discovery --- accelerating the path 
from hypothesis to optimized material across a wide range of technologically important 
material classes.

\section{Data and Code Availability}
The Cu-alloy and glass-forming alloy datasets are publicly available from the repositories 
reported in their respective references~\cite{gorsse2023dataset, Bobadilla2025}. The LTC 
dataset will be made available upon reasonable request to the corresponding author. Source 
code for reproducing all experiments will be released openly at the project's GitHub 
repository upon publication.

\section{Contribution}
Conceptualization, JF.H.; methodology,JF.H, J.H., R.D.,  Y.F., C.W.; software, JF.H., J.H., R.D., Y.F.; resources, J.H.; writing--original draft preparation, JF.H., J.H., R.D., Y.F., C.W.; writing--review and editing,  J.H; visualization, J.H., Y.F., and R.D.; supervision, J.H.;  funding acquisition, J.H.

\bibliographystyle{unsrt}  
\bibliography{references}

\end{document}